\documentclass[twocolumn,superscriptaddress,prl,final,showpacs,preprintnumbers,amsmath,amssymb]{revtex4}

\usepackage{graphicx}
\setkeys{Gin}{width=3.4in,keepaspectratio}
\usepackage{dcolumn}
\usepackage{bm}

\begin{document}
\newcommand*{\NCCO}[0]{Nd$_{2-x}$Ce$_{x}$CuO$_{4\pm\delta}$}
\newcommand*{\NCCo}[0]{Nd$_{2-x}$Ce$_{x}$CuO$_4$}
\newcommand*{\NCO}[0]{Nd$_2$CuO$_4$}
\newcommand*{\LSCO}[0]{La$_{2-x}$Sr$_x$CuO$_4 $}
\newcommand*{\LNSCO}[0]{La$_{1.6-x}$Nd$_{0.4}$Sr$_x$CuO$_4$}
\newcommand*{\YBCO}[0]{YBa$_2$Cu$_3$O$_{6+\delta}$}
\newcommand*{\LCO}[0]{La$_2$CuO$_4$}
\newcommand*{\LCZMO}[0]{La$_2$Cu$_{1-x}$(Zn,Mg)$_x$O$_4$}
\newcommand*{\CuIon}[0]{Cu$^{\text{2+}}$}
\newcommand*{\Neel}[0]{N\'{e}el}
\newcommand*{\degC}[0]{\ensuremath{^\circ}C }
\newcommand*{\etal}[0]{\textit{et~al.}}

\title{Spin Correlations and Magnetic Order in Nonsuperconducting \NCCO}

\author{ P.K. Mang }
\affiliation{Department of Applied Physics, Stanford University,
Stanford, California 94305}

\author{O.P. Vajk}
\altaffiliation {Present Address: NIST Center for Neutron Research, National
Institute of Standards and Technology, Gaithersburg, Maryland 20899, USA}
\affiliation{Department of Physics, Stanford University, Stanford, California
94305, USA}

\author{A. Arvanitaki}
\affiliation{Department of Physics, Stanford University, Stanford, California
94305, USA}

\author{J.W. Lynn}
\affiliation{NIST Center for Neutron Research, National Institute of Standards
and Technology, Gaithersburg, Maryland 20899, USA}

\author{M. Greven}
\affiliation{Department of Applied Physics, Stanford University, Stanford,
California 94305, USA} \affiliation{Stanford Synchrotron Radiation Laboratory,
Stanford, CA 94309, USA}

\date{\today}

\begin{abstract}
We report quantitative neutron scattering measurements of the
evolution with doping of the N\'eel temperature, the
antiferromagnetic correlations, and the ordered moment of
as-grown, non-superconducting \NCCO~($0 \le x \le 0.18$). The
instantaneous correlation length can be effectively described by
our quantum Monte Carlo calculations for the randomly site-diluted
nearest-neighbor spin-1/2 square-lattice Heisenberg
antiferromagnet. However, quantum fluctuations have a stronger
effect on the ordered moment, which decreases more rapidly than
for the quenched-disorder model.

\end{abstract}

\maketitle

Over the past eighteen years much has been learned about the evolution with
doping of magnetic correlations in the high-temperature superconductors. For
the prototypical material \LSCO~(LSCO), neutron scattering has led to the
observation of the loss of long-range antiferromagnetic order and the
concomitant decrease of two-dimensional (2D) spin correlations
\cite{LSCO:Keimer:Longprb}, an incommensurate spin-density wave
\cite{yamada98}, and coupled charge and magnetic ``stripe" order upon codoping
with Nd \cite{tranquada95}. The richness of the observed behavior is
attributable to the evolution of the cuprate superconductors from
Mott-insulating parent materials which are found to be excellent realizations,
at low energies and long wavelengths, of the spin-1/2 square-lattice Heisenberg
antiferromagnet (SLHAF) \cite{birgeneau99}. Although the vast majority of
research has focused on the hole-doped systems, the electron-doped materials,
typified by \NCCO~(NCCO) \cite{NCCO:Discovery}, have presented an important
challenge to the field of high-$T_{\rm c}$ research and particularly to the
paradigm established from experiments on LSCO. Unlike for LSCO, the magnetic
phase of NCCO is very robust with respect to doping
\cite{luke90,thurston90,matsuda92} and the magnetic correlations remain
commensurate \cite{matsuda92,NCCO:Yamada:Gap}.

Since superconductivity in the high-$T_{\rm c}$ cuprates appears
in close proximity to antiferromagnetic phases, it is essential to
understand the nature of nearby magnetic ground states. As-grown,
NCCO is not superconducting, and the antiferromagnetic phase
extends to the highest concentration for which samples can be
produced ($x\approx 0.18$). Oxygen impurities are believed to
occupy the nominally vacant apical sites \cite{schultz96}, and a
relatively severe oxygen reduction procedure must be applied to
induce superconductivity above $x=0.13$ \cite{NCCO:Discovery}.
Although the reduction procedure weakens the magnetism, magnetic
order is still observed near optimal doping where $T_{\rm c} = 24$
K \cite{matsuda92,uefuji01}. One qualitative argument often employed to
explain the robustness of the antiferromagnetic phase in the
electron-doped cuprates is that, whereas hole carriers are doped
into the O 2p band and frustrate the antiferromagnetic arrangement
of the neighboring Cu ions, doped electrons primarily reside on
the Cu site, filling the 3d shell and removing the spin degree of
freedom \cite{luke90,thurston90,matsuda92}. New, quantitative
insights have been gained recently into the role of $static$
non-magnetic impurities introduced into the spin-1/2 SLHAF
\cite{sachdev99,LCZMO:Sandvik:Moment,vajk:experimental,bilayer},
and it would be interesting to test how well NCCO can be described
by this model.

In this Letter, we present detailed neutron scattering results on
the evolution with Ce doping of the magnetic properties of
as-grown, non-superconducting \NCCO~($0 \le x \le 0.18$). We find
that the instantaneous spin correlations in the paramagnetic phase
are, in effect, well described by the randomly-diluted spin-1/2
nearest-neighbor (NN) SLHAF model for which we have carried out
numerical simulations. On the other hand, the ordered moment falls
off more rapidly, a clear indication that quantum fluctuations are
stronger than for the model magnet and that such fluctuations
manifest themselves differently for different observables. We also
find that the primary difference between the N\'eel temperature
and ordered moment of as-grown and reduced NCCO is a shift in
electron concentration that corresponds to $|\Delta x| \approx
0.03$.

Crystals of \NCCO~were grown in 4 atm of O$_2$ using the traveling-solvent
floating-zone technique. The solubility limit is about $x=0.18$, and our study
spans the range $0 \le x \le 0.18$. In addition, several samples were reduced
for 20 h at 960\degC in Ar, followed by an anneal for 20 h at 500\degC in
O$_2$, for additional magnetic order parameter measurements. Magnetometry
indicates that the reduced $x=0.14$ and $x=0.18$ samples are superconducting
with $T_{\rm c} = 24$ and 20 K (onset), respectively. Our single-grain crystals
are cylindrical, typically with a diameter of 4.5 mm, and range in volume from
0.1 to 0.7 cm$^3$. Average Ce concentrations were verified using inductively
coupled plasma spectroscopy and typically found to equal the nominal values.
The neutron scattering measurements were performed on the thermal triple-axis
instruments BT2 and BT9 of the National Institute of Standards and Technology
Center for Neutron Research.

\begin{figure}

  \includegraphics{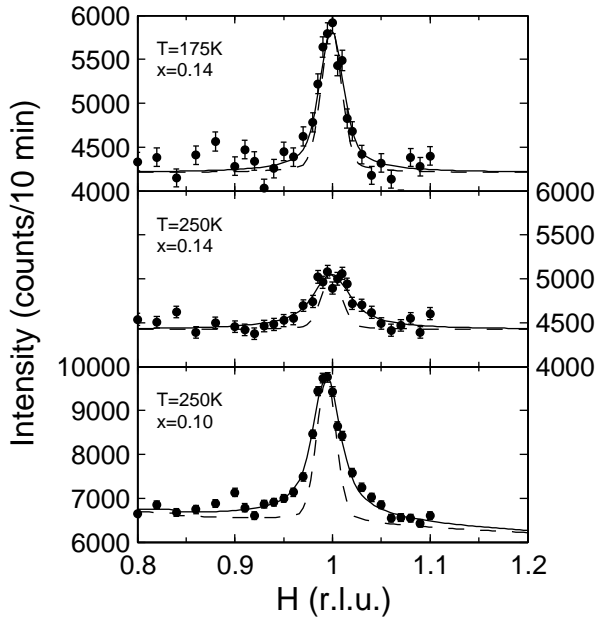}\\

 \vspace{-3mm}

\caption{Representative two-axis scans for $x=0.10$ and 0.14 along
${\bf q}_{\rm 2D} = (H/2 - 1/2, H/2 - 1/2)$ in the paramagnetic
phase with collimations $40^\prime$-$42.5^\prime$-sample-23.7$^\prime$ and incident
neutron energy $E_{\rm i} = 14.7$ meV. The instrumental resolution
is indicated by the dashed lines.}

\end{figure}

All as-grown samples exhibit long-range magnetic order at low temperature.
However, significant two-dimensional (2D) magnetic correlations are already
present well above the \Neel~temperature \cite{matsuda92}. These may be probed
through ``two-axis" scans in which one integrates the dynamic structure factor
$S(\textbf{q}_{\rm 2D},\omega)$ over the limits set by the thermal energy,
$-k_\text{B}T$, and the incident neutron energy, $E_i$. If the integration is
carried out over the relevant fluctuations one obtains the equal-time structure
factor $S(\textbf{q}_{\rm 2D})$ \cite{birgeneau99}. In Fig. 1, we present
representative scans of this type that show a peak in $S(\textbf{q}_{\rm 2D})$
at the incipient antiferromagnetic ordering wavevector. We fit these data to an
isotropic 2D Lorentzian, $S(q_{\rm 2D}) = S(0)/(1+q_{\rm 2D}^2\xi^2)$,
convolved with the instrumental resolution. As the temperature or the doping
level is increased, the correlation length $\xi$ decreases, which manifests
itself as a broadening of the peaks.

\begin{figure}

  \includegraphics{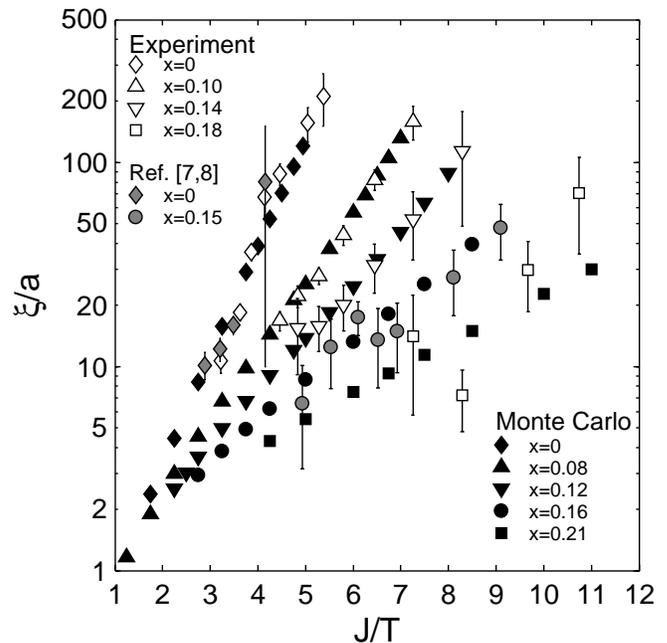}\\

  \vspace{-3mm}
  \caption{Semi-log plot of the magnetic correlation length (in
units of the lattice constant $a$) versus inverse temperature (in
units of  $J = 125$ meV $\approx 1450$ K) for $x=0, 0.10, 0.14,$ and 0.18. The data
terminate just above the onset of N\'eel order at low temperature.
Also displayed are previous data for $x=0$ and $x=0.15$
\cite{thurston90,matsuda92} and Quantum Monte Carlo results for Eq. (1).  }

\end{figure}

Figure 2 summarizes the behavior of $\xi(x,T)$ for several Ce concentrations.
Over a wide temperature range, the spin correlations exhibit an exponential
dependence on inverse temperature, $\xi(x,T) \sim \exp{[2\pi\rho_s(x)/T]}$ with
spin stiffness $\rho_{\rm s} (x)$, indicative of an underlying ground state
with 2D antiferromagnetic order \cite{chakravarty89}. We overlay the results of
quantum Monte Carlo simulations for the randomly-diluted SLHAF, described by
the Hamiltonian
\begin{equation}
{\cal H} = J\sum_{\langle i,j\rangle}{p_ip_j{\bf S}_i\cdot {\bf S}_j},
\end{equation}
where the sum is over NN sites, $J$ is the antiferromagnetic Cu-O-Cu
superexchange, ${\bf S}_i$ is the spin-1/2 operator at site {\it i}, $p_i=1$ on
magnetic sites, and $p_i=0$ on nonmagnetic sites. The numerical method is the
same as that employed previously \cite{vajk:experimental}. The data are plotted
as $\xi/a$, the ratio of the spin-spin correlation length to the NN Cu-Cu
distance ($a\approx3.95$ \AA), versus $J/T$, the ratio of the NN
antiferromagnetic superexchange of \NCO~to the temperature. A previous estimate
from comparison of $\xi(T)$ with theory for the SLHAF yielded $J\approx130$ meV
\cite{thurston90}, and we find $J = 125(4)$ meV from comparison with our
numerical results for $x=0$. Since the value of $J$ has been established, the
only adjustable parameter in the comparison for the Ce-doped samples is the
concentration of nonmagnetic sites in Eq. (1). We find good qualitative
agreement between experiment and Monte Carlo if we fix this concentration to
equal the nominal Ce concentration $x$. However, as shown in Fig. 2, the
agreement becomes quantitative if we allow the effective dilution to differ
slightly: $x_{\rm eff}(x=0.10) = 0.08(1)$, $x_{\rm eff}(x=0.14) = 0.12(1)$, and
$x_{\rm eff}(x=0.18) = 0.21(2)$. Also shown are previous data for as-grown
samples with $x=0$ and $x=0.15$ \cite{thurston90,matsuda92}. The result for
\NCO~agrees well with ours, and we estimate that $x_{\rm eff} (x=0.15) =
0.16(2)$. The extent to which the quenched-disorder model Eq. (1) effectively
describes the spin correlations of as-grown, non-superconducting NCCO is
remarkable. The observed small differences between nominal electron
concentration and effective static dilution may be in part due to the
uncertainty in the oxygen (and hence electron) concentration of our samples.
Another likely possibility is the presence of additional quantum fluctuations
which are not captured in the effective Hamiltonian Eq. (1), for example due to
frustrating next-NN interactions.

The spin stiffness of the disorder-free spin-1/2 NN SLHAF is known
to be $2\pi\rho_{\rm s}/J = 1.131(3)$ \cite{beard98}. An estimate
for the case of quenched disorder, Eq. (1), has been obtained in
previous numerical work \cite{vajk:experimental}. Comparison of
the experimental results in Fig. 2 gives the effective values
$2\pi\rho_{\rm s}/J = 0.71(2)$, 0.54(4), 0.40(7), and 0.25(5),
respectively, for $x = 0.10$, 0.14, 0.15, and 0.18. The value
0.40(7) for the $x=0.15$ sample is somewhat smaller than the
original estimate of $2\pi\rho_{\rm s}/J \approx 0.54$
\cite{matsuda92} which was based on a classical rather than a
quantum \cite{vajk:experimental} picture for the doping dependence
of the spin stiffness.

We note that Ref. \cite{matsuda92} reports data for two as-grown
$x=0.15$ samples with onset of magnetic order at $T_{\rm N} = 160$
and 125 K. If we allow for a distribution of N\'eel temperatures
to account for the rounding of the observed transition, we
estimate mean values of $T_{\rm N} = 147(17)$ and 115(9) K,
respectively. This is consistent with how we have treated our own
data. The second sample from Ref. \cite{matsuda92} also exhibits
lower values of $\xi (T)$ (not shown in Fig. 2), comparable to our
$x=0.18$ crystal. The removal of oxygens through the reduction
process changes both the degree of disorder and the carrier
concentration. This lowers the N\'eel temperature and, for $x >
0.13$, leads to a decrease of the antiferromagnetic volume
fraction and to bulk superconductivity \cite{uefuji01}. Growth at
lower oxygen partial pressure mimics the reduction process,
resulting in lower oxygen contents and higher effective doping
levels. The primary focus of this Letter is on
non-superconducting, as-grown NCCO, and therefore we were careful
to grow our samples under high oxygen partial pressure, resulting
in crystals with $T_{\rm N}$ at or close to the maximum value
attainable.

Figure 3(a) indicates that the N\'eel temperature $T_{\rm N} (x)$
of our as-grown samples, as obtained from neutron diffraction,
approximately follows a parabolic form, extrapolating to zero at
$x \approx 0.21$. As shown in Fig. 3(a), qualitatively similar
behavior was found previously for reduced NCCO for which
$T_{\rm N} (x)$ is lower \cite{uefuji02}. The primary effect of
the reduction is an approximately rigid shift by $|\Delta x|
\approx 0.03$ due to the change in carrier density. The concave
doping dependence for NCCO is contrasted by the behavior exhibited
by the randomly-diluted spin-1/2 SLHAF \LCZMO~(LCZMO), for which N\'eel order extends up to the site
percolation threshold, $x_{\rm p} \approx 0.41,$ and the 2D spin
correlations are quantitatively described by Eq. (1) up to $x_{\rm
p}$ \cite{vajk:experimental}.

\begin{figure}
  \includegraphics{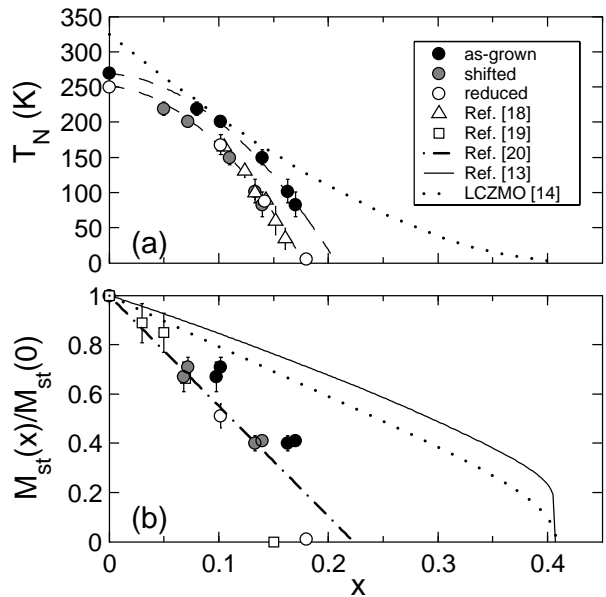}\\
  \vspace{-3mm}
  \caption{(a) \Neel~temperature doping dependence. Ce-doped samples exhibit
some rounding in the transition due to oxygen and cerium inhomogeneities, which
is represented by the error bars. Results from a previous study \cite{uefuji02}
of reduced samples are shown for comparison (triangles). The dashed curves are
quadratic fits, and the dotted line indicates the behavior observed for LCZMO
\cite{vajk:experimental}. The grey circles are the as-grown data shifted by
$|\Delta x| = 0.03$. (b) Doping dependence of the ordered copper moment per
site, normalized by the value for as-grown \NCO. The grey circles are the
as-grown data shifted by $|\Delta x| = 0.03$. The squares are data obtained in
a previous study \cite{Rosseinsky91} and the dot-dashed line is a recent
theoretical prediction \cite{markiewicz03,kusko02,bobm03}. For comparison,
Monte Carlo results for the randomly diluted NN spin-1/2 SLHAF (solid line)
\cite{LCZMO:Sandvik:Moment} and the behavior observed for LCZMO (dotted line)
\cite{vajk:experimental}
are indicated as well.}
\end{figure}

In Fig. 3(b), we plot the ordered moment for several samples, as
obtained from order parameter measurements at low temperature. The
values are normalized to that of \NCO. Consistent with the
behavior of $T_{\rm N} (x)$, the ordered moment of as-grown NCCO
approaches zero much more rapidly than in the case of random
dilution: at the rate $1 - M_{\rm st} (x)/M_{\rm st} (0) \approx
3.5x$ which is approximately twice that of the diluted spin-1/2 SLHAF for
$x < 0.20.$ On the other hand, the magnetic phase extends much
further than for LSCO \cite{LSCO:Keimer:Longprb} and Li-doped
\LCO~\cite{sasagawa02}, for which \Neel~order is destroyed above
$x \approx 0.02$ and $x \approx 0.03$, respectively. The ordered moment of
reduced samples is lowered further. We note that these data are
again normalized by the value for as-grown \NCO, and that a rigid
shift of $|\Delta x| = 0.03$ leads to a good agreement of the two
data sets. Our result is consistent with previous experimental
work at lower concentrations \cite{Rosseinsky91}.

If viewed instantaneously, so as to mitigate the effects due to the increased
itinerancy upon doping, the magnetism of NCCO should indeed resemble a system
with quenched random site dilution. What is surprising is that for $\xi(x,T)$
the correspondence with Eq. (1) is nearly quantitative. Since the charge
carriers in as-grown NCCO are itinerant, even well below $T_{\rm N}$
\cite{transport}, deviations from the static model Eq. (1) may be expected to
emerge as we view the two systems on different time scales. Indeed, the ordered
moment, a measure of the strength of the magnetic order at infinitely long
times, decreases much more rapidly in the case of NCCO than for the SLHAF. We
note that slight deviations, possibly due to the presence of a frustrating
next-NN exchange, are already observable for the ordered moment of LCZMO
\cite{vajk:experimental} [Fig. 3(b)], and that LCZMO appears to be in close
proximity to a new quantum critical point in an extended parameter space
\cite{bilayer}. In the case of LCZMO, these perturbations are not quite strong
enough to shift the critical point away from the geometric site-percolation
threshold. Although the experimentally accessible doping range is limited, our
results for NCCO are consistent with the existence of a quantum critical point
below the geometric site-percolation threshold.

We note that recent calculations for the one-band Hubbard model are in
semi-quantitative agreement with our data for the spin correlations
\cite{markiewicz03,kyung03} and ordered moment
\cite{markiewicz03,kusko02,bobm03}. The latter result is shown in Fig. 3(b).
These theories were motivated by photoemission measurements of the evolution of
the Fermi surface of reduced \NCCO~($0\le x \le 0.15$) \cite{armitage02}. Their
proponents assert that \textit{t\textendash J} models are intrinsically limited
in describing the electron-doped cuprates and that a proper description
requires \textit{t\textendash U} models with a decreasing value of $U$ with
increasing electron concentration
\cite{kusko02,markiewicz03,senechal03,kyung03}.

In summary, we have presented a comprehensive neutron scattering
study of the magnetic properties of non-superconducting \NCCO.
Although the evolution of the instantaneous spin-correlations may
effectively be described by a site-dilution model, a proper
description of the magnetic degrees of freedom must necessarily
include the effects of electron itinerancy. While the achievements
of recent theoretical treatments of the Hubbard model seem very
promising, it is not clear at present if this model exhibits a
superconducting ground state. Phonon anomalies have been observed
in both electron and hole doped high-temperature superconductors
\cite{dastuto02}, and a complete description of these materials
may require the additional consideration of the electron-lattice
coupling. Regardless, our data should serve as a benchmark for
tests of still-emerging theories for the high-$T_{\rm c}$ phase
diagram.

We acknowledge valuable discussions with N.P. Armitage, S.
Larochelle, R.S. Markiewicz, M. Matsuda, and A.-M. S. Tremblay,
and thank I. Tanaka and the late H. Kojima for their invaluable
assistance during the initial stage of the crystal growth effort.
This work was supported by the US Department of Energy under
Contracts No. DE-FG03-99ER45773 and No. DE-AC03-76SF00515, and by
NSF CAREER Award No. DMR9985067.

\vspace{-5mm}

\bibliography{MangPRL} 

\begin{thebibliography}{28}
\expandafter\ifx\csname natexlab\endcsname\relax\def\natexlab#1{#1}\fi
\expandafter\ifx\csname bibnamefont\endcsname\relax
  \def\bibnamefont#1{#1}\fi
\expandafter\ifx\csname bibfnamefont\endcsname\relax
  \def\bibfnamefont#1{#1}\fi
\expandafter\ifx\csname citenamefont\endcsname\relax
  \def\citenamefont#1{#1}\fi
\expandafter\ifx\csname url\endcsname\relax
  \def\url#1{\texttt{#1}}\fi
\expandafter\ifx\csname urlprefix\endcsname\relax\def\urlprefix{URL }\fi
\providecommand{\bibinfo}[2]{#2}
\providecommand{\eprint}[2][]{\url{#2}}

\bibitem[{LSC()}]{LSCO:Keimer:Longprb}
\bibinfo{note}{B. Keimer {\it et al.}, \prb {\bf 46}, 14034 (1992).}

\bibitem[{yam()}]{yamada98}
\bibinfo{note}{K. Yamada {\it et al.}, \prb {\bf 57}, 6165 (1998), and
  references therein.}

\bibitem[{tra({\natexlab{a}})}]{tranquada95}
\bibinfo{note}{J.~M. Tranquada {\it et al.}, Nature (London) {\bf 375}, 561
  (1995).}

\bibitem[{bir()}]{birgeneau99}
\bibinfo{note}{R.~J. Birgeneau {\it et al.}, \prb {\bf 59}, 13788 (1999).}

\bibitem[{\citenamefont{Tokura et~al.}(1989)\citenamefont{Tokura, Takagi, and
  Uchida}}]{NCCO:Discovery}
\bibinfo{author}{\bibfnamefont{Y.}~\bibnamefont{Tokura}},
  \bibinfo{author}{\bibfnamefont{H.}~\bibnamefont{Takagi}}, \bibnamefont{and}
  \bibinfo{author}{\bibfnamefont{S.}~\bibnamefont{Uchida}},
  \bibinfo{journal}{Nature} \textbf{\bibinfo{volume}{337}},
  \bibinfo{pages}{345} (\bibinfo{year}{1989}).

\bibitem[{luk()}]{luke90}
\bibinfo{note}{G.~M. Luke {\it et al.}, \prb {\bf 42}, 7981 (1990).}

\bibitem[{thu()}]{thurston90}
\bibinfo{note}{T.~R. Thurston {\it et al.}, \prl {\bf 65}, 263 (1990).}

\bibitem[{mat()}]{matsuda92}
\bibinfo{note}{M. Matsuda {\it et al.}, \prb {\bf 45}, 12548 (1992).}

\bibitem[{\citenamefont{Yamada et~al.}(1999)\citenamefont{Yamada, Kurahashi,
  Endoh, Birgeneau, and Shirane}}]{NCCO:Yamada:Gap}
\bibinfo{author}{\bibfnamefont{K.}~\bibnamefont{Yamada}},
  \bibinfo{author}{\bibfnamefont{K.}~\bibnamefont{Kurahashi}},
  \bibinfo{author}{\bibfnamefont{Y.}~\bibnamefont{Endoh}},
  \bibinfo{author}{\bibfnamefont{R.~J.} \bibnamefont{Birgeneau}},
  \bibnamefont{and} \bibinfo{author}{\bibfnamefont{G.}~\bibnamefont{Shirane}},
  \bibinfo{journal}{J. Phys. Chem. Solids} \textbf{\bibinfo{volume}{60}},
  \bibinfo{pages}{1025} (\bibinfo{year}{1999}).

\bibitem[{\citenamefont{Schultz et~al.}(1996)\citenamefont{Schultz, Jorgensen,
  Peng, and Greene}}]{schultz96}
\bibinfo{author}{\bibfnamefont{A.~J.} \bibnamefont{Schultz}},
  \bibinfo{author}{\bibfnamefont{J.~D.} \bibnamefont{Jorgensen}},
  \bibinfo{author}{\bibfnamefont{J.~L.} \bibnamefont{Peng}}, \bibnamefont{and}
  \bibinfo{author}{\bibfnamefont{R.~L.} \bibnamefont{Greene}},
  \bibinfo{journal}{\prb} \textbf{\bibinfo{volume}{53}}, \bibinfo{pages}{5157}
  (\bibinfo{year}{1996}).

\bibitem[{uef()}]{uefuji01}
\bibinfo{note}{T. Uefuji {\it et al.}, Physica (Amsterdam) {\bf 357C}, 208
  (2001).}

\bibitem[{\citenamefont{Sachdev et~al.}(1999)\citenamefont{Sachdev, Buragohain,
  and Vojta}}]{sachdev99}
\bibinfo{author}{\bibfnamefont{S.}~\bibnamefont{Sachdev}},
  \bibinfo{author}{\bibfnamefont{C.}~\bibnamefont{Buragohain}},
  \bibnamefont{and} \bibinfo{author}{\bibfnamefont{M.}~\bibnamefont{Vojta}},
  \bibinfo{journal}{Science} \textbf{\bibinfo{volume}{286}},
  \bibinfo{pages}{2479} (\bibinfo{year}{1999}).

\bibitem[{\citenamefont{Sandvik}(2002)}]{LCZMO:Sandvik:Moment}
\bibinfo{author}{\bibfnamefont{A.~W.} \bibnamefont{Sandvik}},
  \bibinfo{journal}{\prb} \textbf{\bibinfo{volume}{66}},
  \bibinfo{pages}{024418} (\bibinfo{year}{2002}).

\bibitem[{vaj()}]{vajk:experimental}
\bibinfo{note}{O. P. Vajk {\it et al.}, Science {\bf 295}, 1691 (2002); O. P.
  Vajk, M. Greven, P. K. Mang, and J. W. Lynn, Solid State Comm. {\bf 126}, 93
  (2003)}.

\bibitem[{bil()}]{bilayer}
\bibinfo{note}{A. W. Sandvik, \prl {\bf 89}, 177201 (2002); O. P. Vajk and M.
  Greven, \prl {\bf 89}, 172202 (2002)}.

\bibitem[{\citenamefont{Chakravarty et~al.}(1989)\citenamefont{Chakravarty,
  Halperin, and Nelson}}]{chakravarty89}
\bibinfo{author}{\bibfnamefont{S.}~\bibnamefont{Chakravarty}},
  \bibinfo{author}{\bibfnamefont{B.~I.} \bibnamefont{Halperin}},
  \bibnamefont{and} \bibinfo{author}{\bibfnamefont{D.~R.}
  \bibnamefont{Nelson}}, \bibinfo{journal}{\prb} \textbf{\bibinfo{volume}{39}},
  \bibinfo{pages}{2344} (\bibinfo{year}{1989}).

\bibitem[{\citenamefont{Beard et~al.}(1998)\citenamefont{Beard, Birgeneau,
  Greven, and Wiese}}]{beard98}
\bibinfo{author}{\bibfnamefont{B.~B.} \bibnamefont{Beard}},
  \bibinfo{author}{\bibfnamefont{R.~J.} \bibnamefont{Birgeneau}},
  \bibinfo{author}{\bibfnamefont{M.}~\bibnamefont{Greven}}, \bibnamefont{and}
  \bibinfo{author}{\bibfnamefont{U.-J.} \bibnamefont{Wiese}},
  \bibinfo{journal}{\prl} \textbf{\bibinfo{volume}{80}}, \bibinfo{pages}{1742}
  (\bibinfo{year}{1998}).

\bibitem[{\citenamefont{Uefuji et~al.}(2002)}]{uefuji02}
\bibinfo{author}{\bibfnamefont{T.}~\bibnamefont{Uefuji}} \bibnamefont{et~al.},
  \bibinfo{journal}{Physica (Amsterdam)} \textbf{\bibinfo{volume}{378C-381C}},
  \bibinfo{pages}{273} (\bibinfo{year}{2002}).

\bibitem[{\citenamefont{Rosseinsky et~al.}(1991)\citenamefont{Rosseinsky,
  Prassides, and Day}}]{Rosseinsky91}
\bibinfo{author}{\bibfnamefont{M.~J.} \bibnamefont{Rosseinsky}},
  \bibinfo{author}{\bibfnamefont{K.}~\bibnamefont{Prassides}},
  \bibnamefont{and} \bibinfo{author}{\bibfnamefont{P.}~\bibnamefont{Day}},
  \bibinfo{journal}{Inorg. Chem.} \textbf{\bibinfo{volume}{30}},
  \bibinfo{pages}{2680} (\bibinfo{year}{1991}).

\bibitem[{mar()}]{markiewicz03}
\bibinfo{note}{R.~S. Markiewicz, cond-mat/0312594.}

\bibitem[{kus()}]{kusko02}
\bibinfo{note}{C. Kusko, R.~S. Markiewicz, M. Lindroos and A. Bansil, Phys.
  Rev. B {\bf66}, 140513R (2002).}

\bibitem[{bob()}]{bobm03}
\bibinfo{note}{Ref. \cite{kusko02} is a mean-field theory. Up to $x=0.10$,
  quantum fluctuations lower $M_{\rm st} (x)$ by $1/\eta \sim 0.78 - 0.84$, so
  the ratio $M_{\rm st} (x)$/$M_{\rm st} (0)$ remains approximately unaltered.
  The linear behavior at higher concentrations is to be viewed as an upper
  bound \cite{markiewicz03}.}

\bibitem[{sas()}]{sasagawa02}
\bibinfo{note}{T. Sasagawa {\it et al.}, \prb {\bf 66}, 184512 (2002).}

\bibitem[{tra({\natexlab{b}})}]{transport}
\bibinfo{note}{The planar resistivity at 4.2 K is as low as $\rho \sim 2$
  m$\Omega$ cm for as-grown NCCO ($x=0.15$): Y. Onose {\it et al.}, \prl {\bf
  82}, 5120 (1999).}

\bibitem[{kyu()}]{kyung03}
\bibinfo{note}{B. Kyung, V. Hankevych, A.-M. Dar\'{e}, and A.-M.~S. Tremblay,
  cond-mat/0312499.}

\bibitem[{arm()}]{armitage02}
\bibinfo{note}{N.~P. Armitage {\it et al.}, \prl {\bf 88}, 257001 (2002).}

\bibitem[{sen()}]{senechal03}
\bibinfo{note}{D. S\'{e}n\'{e}chal and A.-M.~S. Tremblay, \prl {\bf 92}, 126401
  (2004).}

\bibitem[{das()}]{dastuto02}
\bibinfo{note}{M. d'Astuto {\it et al.}, \prl {\bf 88}, 167002 (2002), and
  references therein.}

\end{thebibliography}

\end{document}